\documentclass[aps,prl,twocolumn, showpacs,amsmath,amssymb,superscriptaddress]{revtex4}
\usepackage{graphicx}
\usepackage{dcolumn}
\usepackage{bm}
\usepackage{amsmath}
\usepackage{amssymb}
\usepackage{epsfig}
\begin{document}
\title{Phase-Dependent Electronic Specific Heat in Mesoscopic Josephson Junctions}
\author{Hassan Rabani}
\email{h_rabani@sci.ui.ac.ir}
\affiliation{Department of Physics, Faculty of Sciences, University of Isfahan, 81744 Isfahan, Iran}
\affiliation{NEST CNR-INFM and Scuola Normale Superiore, I-56126 Pisa, Italy}
\author{Fabio Taddei}
\affiliation{NEST CNR-INFM and Scuola Normale Superiore, I-56126 Pisa, Italy}
\author{Olivier Bourgeois}
\affiliation{Institut N\`eel, CNRS-UJF, 25 av des Martyrs, BP 166, 38042 Grenoble Cedex 9, France}
\author{Rosario Fazio}
\affiliation{International School for Advanced Studies (SISSA),  Via Beirut 2-4, I-34014 Trieste, Italy}
\affiliation{NEST CNR-INFM and Scuola Normale Superiore, I-56126 Pisa, Italy}
\author{Francesco Giazotto}
\email{giazotto@sns.it}
\affiliation{NEST CNR-INFM and Scuola Normale Superiore, I-56126 Pisa, Italy}
\begin{abstract}
We study the influence of superconducting correlations on the electronic specific heat in a diffusive superconductor-normal metal-superconductor Josephson junction. We present a description of this system in the framework of the diffusive-limit Green's function theory, taking into account finite temperatures, phase difference as well as junction parameters.  We find that \emph{proximity effect} may lead to a substantial deviation of the specific heat as compared to that in the normal state, and that it can be largely tuned in magnitude  by changing the phase difference between the superconductors. A measurement setup to confirm these predictions is also suggested.
\end{abstract}
\maketitle

Superconducting circuits  attract much interest thanks to their potential in applications such as single-electron devices \cite{pekola,saira}  and Josephson junctions \cite{c5}, as well as for radiation detection \cite{day}. 
In this context, \emph{proximity effect} \cite{c1} in mesoscopic normal metal-superconductor nanostructures has drawn researchers attention \cite{taddei,recent}. In such systems the normal metal can acquire superconducting-like properties which manifest themselves in macroscopic observables, such as the Josephson current \cite{c5}.
Although the focus has been mainly given to the electronic properties of  proximity structures, thermal transport is recently under the spotlight
\cite{rmp}.  
Yet, very little is known about  the thermodynamic quantities of these systems, for instance, the entropy as well as the specific heat,  which could be crucial for future applications in nanoelectronics.   

In this Letter we theoretically address the influence of superconducting correlations on the electronic specific heat of a diffusive superconductor-normal metal-superconductor (SNS) Josephson junction. We show that, thanks to proximity effect, the specific heat of the N region can be dramatically different from that in the absence of superconductivity.  In particular, at low temperature, it is substantially suppressed with respect to the normal state, and turns out to be largely tunable in magnitude by changing the phase difference between the S reservoirs.
Such peculiarity of  mesoscopic SNS Josephson junctions may have impact for the implementation of novel devices, for instance,  ultrasensitive single-photon detectors based on proximity effect \cite{pjs}.
\begin{figure}[t]	
		\includegraphics[width=\columnwidth,clip]{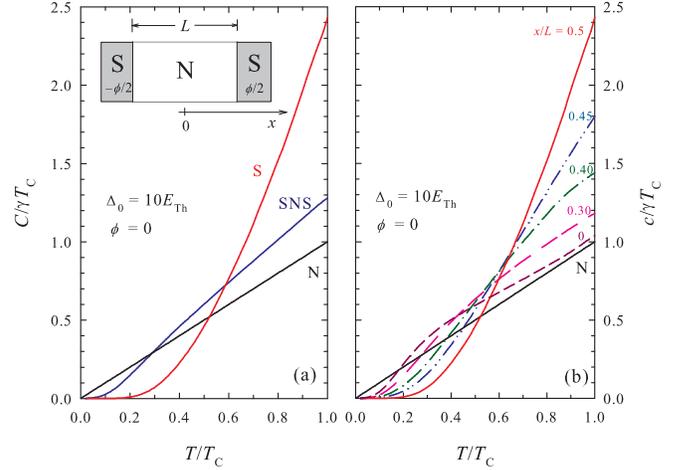}
	\caption{(color online) (a) Comparison of the SNS electronic specific heat $C$ vs temperature $T$  with that in the normal state (N) and in the superconductor (S) calculated for $\Delta_0=10E_{{\rm Th}}$ and $\phi=0$. The inset shows a scheme of the SNS junction of length $L$. $x=0$ denotes the middle of the junction, and the system is assumed quasi-one-dimensional. (b) $c$ vs $T$  at different positions  $x$ along the N wire calculated for $\Delta_0=10E_{{\rm Th}}$ and $\phi=0$. Also shown is the specific heat in the absence of superconductivity (N).
	$T_{\text{C}}$ is the critical temperature of the superconductor.}
	\label{fig1}
\end{figure}

The mesoscopic system under investigation [see the inset of Fig. \ref{fig1}(a)] consists of a diffusive normal metal wire of length $L$ in contact with two superconducting electrodes kept at zero potential, thus defining a SNS Josephson junction. The wire is oriented along the $x$ direction, and  we assume  its transverse dimensions to be much smaller than $L$, so that the wire can be considered as quasi-one-dimensional.
The contact with superconducting leads induces in the N region correlations through proximity effect \cite{c1}, which is responsible for the Josephson supercurrent flow across the structure. 

The proximity effect in the normal metal region can be described with the Usadel equations \cite{usadel}.
According to the parametrization of the Green's functions given in Ref. \cite{c4},
the Usadel equations in the N region can be written as \cite{usadel,c4}
\begin{eqnarray}
\hbar D\partial^2_x\theta =-2iE\sinh(\theta) +\frac{\hbar D}{2}
\left(\partial_x\chi \right)^2\sinh(2\theta) \nonumber\\
\text{sinh}(2\theta)\partial_x \theta\partial_x \chi+\text{sinh}^2(\theta)\partial_x^2\chi=0,
\label{retard}
\end{eqnarray} 
where
$D$  is the diffusion coefficient and the energy $E$ is relative to the chemical potential in the superconductors.
$\theta$ and $\chi$ are, in general, complex scalar functions of position $(x)$ and energy.
For perfectly transmitting interfaces \cite{interface} the boundary conditions at the NS contacts reduce to
\begin{equation}
  \begin{array}[c]{rclcl}
    \theta(\pm L/2)&=&\hbox{arctanh}(\Delta/E)\;,\\
    \chi(\pm L/2)&=&\pm\phi/2\;,
  \end{array}\label{bc}
\end{equation}
where $\phi$ is the phase difference between S electrodes, and $\Delta$ is the superconducting order parameter. 
We adopt a step-function form for the order parameter, i.e., constant in S and zero in the N wire, although $\Delta$ is in principle $x$-dependent and can be  determined self-consistently \cite{c4}. 
Furthermore, we assume the usual BCS temperature dependence of $\Delta(T)$ with critical temperature $T_{\text{C}}=\Delta_0/(1.764k_{\text{B}})$, where $\Delta_0$ is the zero-temperature order parameter, and
$k_{\text{B}}$ is the Boltzmann constant.
The density of states (DOS) in the N region is then given by
\begin{equation}\label{density}
   \mathcal{N}(x,E,T,\phi)=\mathcal{N}_{\text{F}} \mbox{Re}\left\{\cosh\left[\theta(x,E,T,\phi)\right]\right\} \; ,
\end{equation}
where $\mathcal{N}_{\text{F}}$ is the normal-state DOS at the Fermi level. 
Assuming that the S electrodes are not influenced by the proximity effect,
the total electronic specific heat ($C^{\text{tot}}$) of the SNS junction is determined by
\begin{equation}
 C^{\text{tot}}(T,\phi)=C(T,\phi)+C_{\text{S}}(T),
\label{ctot}
\end{equation}
where $C_{\text{S}}(T)$ is the temperature-dependent specific heat of the superconducting leads, while
\begin{equation}
 C(T,\phi)=T\frac{\partial \mathcal{S}(T,\phi)}{\partial T}, 
\end{equation}
is the specific heat of the N region. Furthermore, the entropy $\mathcal{S}(T,\phi)$ is defined as \cite{entropy} 
 \begin{eqnarray}
\mathcal{S}(T,\phi)=-\frac{4}{L}\int^{L/2}_{-L/2}\int^{\infty}_{0} dx dE \mathcal{N}(x,E,T,\phi)\\
\times\left[f(E)\ln(f(E))+(1-f(E))\ln(1-f(E))\right],\nonumber
\label{entropy}
\end{eqnarray}
and $f(E)=(1+\exp[E/(k_{\text{B}}T)])^{-1}$ is the Fermi-Dirac quasiparticle distribution function at temperature $T$.
The characteristic energy scale for the N region is $E_{{\rm Th}}=\hbar D/L^2$, usually referred to as the Thouless energy.
In the following we shall focus on \emph{long} SNS junctions only 
($\Delta_0\gg E_{{\rm Th}}$), as it is the relevant regime  for metallic diffusive Josephson junctions \cite{c7,c6}. 

By numerically solving Eqs. (\ref{retard}) with the above given boundary conditions [see Eqs. (\ref{bc})] we can determine the DOS as a function of energy and position for a given $\phi$ and temperature. In particular, the DOS is an even function of energy, and exhibits a minigap ($E_g$) for  $|E|\leq E_g$ \cite{dos1}  which depends, in general, on $E_{{\rm Th}}$ and $\phi$. In addition, $E_g\approx 3.2 E_{\text{Th}}$ for long junctions at $\phi=0$ \cite{charlat}, and decreases by increasing $\phi$, vanishing at $\phi=\pi$. 
Moreover, the density of states at  NS interfaces (i.e., at $x=\pm L/2$)  equals the BCS superconducting DOS, 
while it tends to $\mathcal{N}_{\text{F}}$ for $ |E|\gg\Delta_0$. 
\begin{figure}[t]
		\includegraphics[width=\columnwidth,clip]{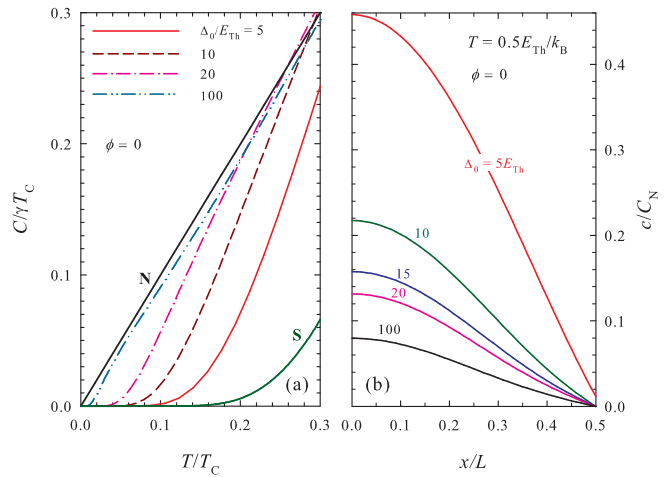}
	\caption{(color online) (a) Specific heat $C$ vs $T$  at  $\phi=0$ calculated for several ratios $\Delta_0/E_{\text{Th}}$. (b) $c/C_{\text{N}}$ vs $x$ at $\phi=0$ and $T=0.5 E_{{\rm Th}}/k_{\text{B}}$ calculated for several ratios $\Delta_0/E_{\text{Th}}$. $C_{\text{N}}=\gamma T$ is the specific heat in the normal state.}
	\label{fig2}
\end{figure}  

In Fig. \ref{fig1}(a) the electronic specific heat $C$ vs temperature $T$ for a superconductor, the N region of a SNS junction, and a normal metal are compared. In the normal metal  case, the  low-temperature specific heat, i.e., $C_{\text{N}}=\gamma T$, where $\gamma=2\pi^2\mathcal{N}_{\text{F}}k_{\text{B}}^2/3$ is plotted. 
For the superconductor, the figure displays the standard BCS result, where $C_{\text{S}}$ follows the low-temperature exponential drop according to $C_{\text{S}}(T)\simeq 1.34 \gamma T_{\text{C}}(\Delta_0/k_{\text{B}}T)^{3/2}e^{-\Delta_0/k_{\text{B}}T}$ for $T \ll T_{\text{C}}$, and exhibits the discontinuous jump $\delta C_{\text{S}}=1.43\gamma T_{\text{C}}$ at the critical temperature \cite{c2}.
In the SNS case we have chosen a moderately-long junction with $\Delta_0=10 E_{{\rm Th}}$, and we set $\phi=0$.
It is  evident that in the N region of a SNS Josephson junction $C$ strongly departs from the normal metal behavior, and resembles that of a superconductor. More in particular, it shows a low-temperature suppression with respect to the N state, with vanishingly small values for $T\ll T_{\text{C}}$, and a discontinuous jump at $T_{\text{C}}$ which is smaller than that in S. The temperature determining the onset of strong suppression is set by the value of the minigap. 

The dependence on position ($x/L$) along the N wire is shown in Fig. \ref{fig1}(b), where the electronic
specific heat per unit length $c(x,T,\phi)$ vs $T$ is displayed for the same SNS junction. 
This quantity is defined according to the expression $C(T,\phi)=\frac{1}{L}\int^{L/2}_{-L/2}dx \,\,c(x,T,\phi)$.
First of all, we note that  at the NS interface (i.e., for $x/L=1/2$) $c$ coincides with the specific heat of a BCS superconductor. 
By moving toward the center of the junction, the discontinuous jump at $T=T_{\text{C}}$ gets reduced, while the temperature which sets the onset of strong suppression becomes smaller. This can be understood by noting that moving  away from the NS interfaces the superconducting correlations induced in the N region become weaker (i.e., the proximity effect gets suppressed), thus making the specific heat to approach that in the normal state.

The role of junction length $L$ is shown in Fig. \ref{fig2}(a), where the SNS specific heat $C$ is plotted as a function of  $T$ for several  $\Delta_0/E_{\text{Th}}$ ratios at $\phi=0$. 
By increasing $L$, or equivalently for large $\Delta_0/E_{\text{Th}}$ ratios, $C$ approaches the normal-state specific heat, which follows from the weakening of the proximity effect in longer SNS junctions. 
Moreover, the temperature which sets the onset of strong suppression decreases by increasing the junction length which stems from a reduction of the proximity-induced minigap in N. 
For instance, for $\Delta_0/E_{\text{Th}}=10$ at $T=0.1 T_{\text{C}}$, the specific heat turns out to be suppressed by about a factor of $5$ as compared to that in the normal state at the same temperature.
This points out that \emph{moderately-long} Josephson junctions are more suitable to maximize the effects of superconducting correlations on the specific heat.

On the other hand, it is interesting to investigate the behavior for fixed junction length, i.e., for fixed $E_{\text{Th}}$, but for superconductors with different energy gap $\Delta_0$. 
Figure \ref{fig2}(b) shows the low-temperature SNS specific heat per unit length vs position $x$ at $T=0.5 E_{\text{Th}}$ and $\phi=0$ for several values of $\Delta_0$. 
By moving away from the middle of the junction ($x=0$) the specific heat strongly decreases, approaching the value of the superconductor at the NS interface ($x=L/2$). 
Furthermore, by reducing the superconducting gap $\Delta_0$, the specific heat value in the middle of the N region tends to that in the normal state. 
For instance, an increase of $\Delta_0$ by a factor of $10$, i.e, from $10E_{\text{Th}}$ to $100E_{\text{Th}}$ which corresponds roughly to change from aluminum (Al) to niobium (Nb) electrodes, translates in a reduction of $C$ by almost a factor of $3$.  
This can be understood  by noting that for superconductors with smaller gap and fixed junction length, the  proximity effect is weakened in the middle of the Josephson junction, leading to an enhancement of the electronic specific heat value.
For a given junction length, larger gap superconductors are thus preferable in order to suppress the low-temperature specific heat.

\begin{figure}[t]
		\includegraphics[width=\columnwidth,clip]{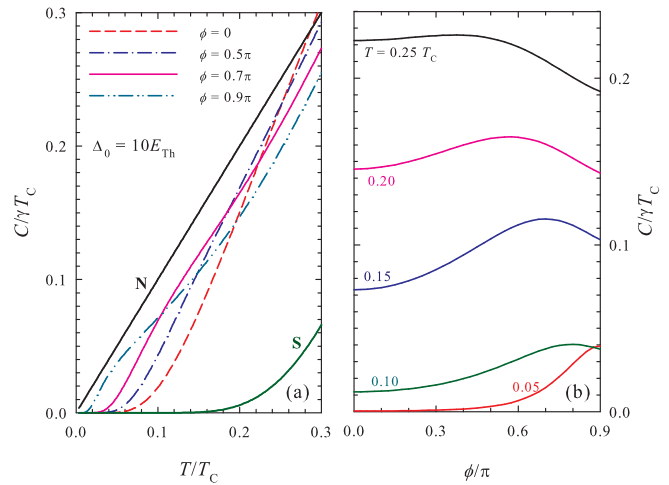}
	\caption{(color online) (a) Specific heat $C$ vs $T$ calculated for  $\Delta_0=10E_{{\rm Th}}$ and different $\phi$ values.  (b) Specific heat $C$ vs $\phi$ calculated for $\Delta_0=10E_{{\rm Th}}$ and several $T$ values.}
	\label{phi}
\end{figure}

Figure \ref{phi}(a) shows the SNS specific heat $C$ versus temperature $T$ for a junction with $\Delta_0=10E_{\text{Th}}$ at different values of the phase difference $\phi$ between the superconductors. 
For a comparison, the electron specific heat in the N and S state are plotted as well.
By increasing $\phi$ from zero, the minigap in the N region DOS decreases \cite{charlat} therefore enhancing the low-temperature specific heat. 
At $\phi=\pi$ (not shown) the minigap vanishes \cite{charlat}, so that $C$ approaches the N state specific heat at very low temperatures (i.e., for $T\rightarrow0$). 

The full phase dependence is plotted in Fig. \ref{phi}(b), which shows the specific heat in the N region of a SNS junction with $\Delta_0=10E_{\text{Th}}$ at different temperatures $T$. 
$C$ is a non-monotonic function of $\phi$, and  it is maximized at an optimal phase difference which strongly depends on $T$. 
In particular, the optimal value shifts toward larger $\phi$ by lowering the temperature. 
Notably, the figure shows that, especially at the lowest temperatures, the specific heat variation with $\phi$ can be extremely large. For instance, at $T=0.1T_{\text{C}}$, the specific heat is enhanced by about $340\%$ from $\phi=0$ to $\phi\simeq 0.8\pi$, while at $T=0.05T_{\text{C}}$ a huge variation of the order of $8800\%$ can be obtained by varying the phase from 0 to $0.9\pi$. 
This demonstrates the sensitivity of the electronic specific heat to $\phi$ which stems from the coherent nature of superconducting correlations induced in SNS Josephson junctions. 

\begin{figure}[t] 
 		\includegraphics[width=\columnwidth,clip]{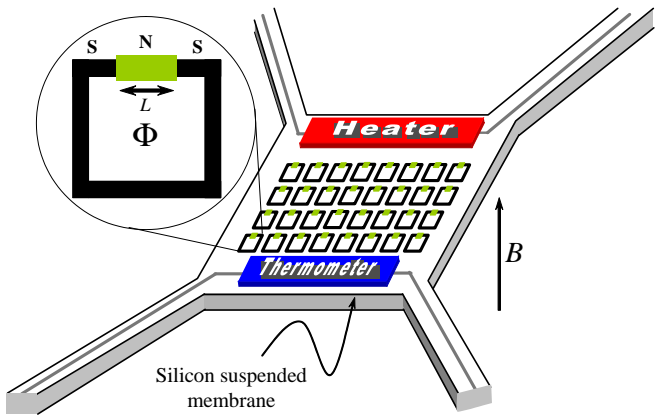}
	\caption{(color online) A possible experimental setup consisting of a SNS ac proximity SQUID threaded by an external magnetic field. 
The loop geometry will allow to change the phase difference at the NS boundaries by the application of a magnetic field ($B$), thus controlling the electronic specific heat of the N region. 
An array consisting of a large number of identical loops ($\sim10^6$) is patterned onto a silicon suspended membrane, while the heater and thermometer will allow to measure the  structure specific heat by ac calorimetry \cite{accal}.}
	\label{exp}
\end{figure}

We now discuss a possible experimental setup for measuring the predicted effects. 
A realistic measurement scheme is shown in Fig. \ref{exp}. 
The basic element of the structure consists of an ac proximity superconducting quantum interference device (SQUID), i.e., a superconducting loop interrupted by a  normal metal region of length $L$. 
The electronic specific heat $C$ in the N region can be controlled by an externally applied magnetic field $B$, giving rise to a total flux $\Phi$ through the loop area. The system can be realized by state-of-the-art \emph{e}-beam lithography and shadow mask evaporation of metallic thin films. An array consisting of a large number of identical proximity SQUIDs  can be deposited on a silicon suspended membrane  on which a thermometer and a heater have been previously patterned. The structure specific heat can be  measured by ac calorimetry \cite{accal}, which consists of inducing temperature oscillations of the thermally isolated membrane and thermometer by supplying ac power to the heater. 

In the following we shall provide some estimates to get the order of magnitude of what can be experimentally measured. 
For our estimate we choose a $50$-nm-thick copper (Cu) wire, with $L=0.6\,\mu$m, width of 200 nm, $D=0.01$ m$^2$s$^{-1}$, and density $\rho=8.92\times10^3$ kgm$^{-3}$. With such parameters we get a N region volume $\mathcal{V}=6\times10^{-21}$ m$^3$, mass $m=5.35\times10^{-17}$ kg, and $E_{\text{Th}}\simeq18\,\mu$eV. 
By choosing Al for the superconducting loop with $\Delta_0=200\,\mu$eV we get $\Delta_0/E_{\text{Th}}\simeq10.9$.
Furthermore, at $T=100$ mK the copper heat capacitance is $\mathcal{C}^{\text{Cu}}_{\text{N}} =2\times10^{-3} $Jkg$^{-1}$K$^{-1}$ which gives a total specific heat  of the N island $C^{\text{Cu}}_{\text{N}} =m\mathcal{C}^{\text{Cu}}_{\text{N}}\simeq 1\times10^{-19}$JK$^{-1}$. 
From our calculations for a SNS Josephson junction with $\Delta_0/E_{\text{Th}}=10$ and $\phi=0$ we have that a specific heat variation of $1600\%$  is predicted by changing the temperature from $0.1T_{\text{C}}$ to $0.2T_{\text{C}}$ (i.e., from about 120 to 240 mK for Al) [see Fig. \ref{fig2}(a)], and a variation of the order of $340\%$ at $T=0.1T_{\text{C}}$ by changing $\phi$ [see Fig. \ref{phi}(b)]. 
Thanks to proximity effect, a change of $C$   of $1.6\times10^{-19}$JK$^{-1}$ by varying the temperature in such a range, and of $3.4\times10^{-20}$JK$^{-1}$ by varying the phase is expected for the system proposed in Fig. \ref{exp}. 
Recent developments in highly-sensitive calorimetry applied to mesoscopic systems allow to measure such small signatures \cite{b1,b2}. 
The sensitivity of these experiments can be estimated of the order of $10k_{\text{B}}$ per $\mu$m$^2$ of the structured silicon membrane. 
A superconducting loop of $1\mu$m of diameter occupies a surface of approximately $4\mu$m$^2$ in order to avoid any interaction with the neighboring loops of the array, so that a minimum of $\sim5.5\times10^{-22}$J/K can be measured. Signal to noise ratios ranging from several tens to some $10^2$
are expected, so that such signatures could be  detected at these temperatures. 
In practice, what will be measured is the specific heat of the whole structure which includes  the contribution from the membrane and transducers, and that related to  $n$ (typically $\sim10^6$) identical loops  of the array. 
The loop specific heat is the sum of the N and S parts according to Eq. (\ref{ctot}).
In particular, at low temperature, i.e., below about $T=0.2T_{\text{C}}$, the N region provides the predominant contribution to the specific heat of the loop [see Fig. \ref{phi}(a)], while at higher temperature the one from the superconductor  becomes relevant. 

In summary, we have presented a microscopic description of the influence of proximity effect on the electronic specific heat of a mesoscopic SNS Josephson junction, and suggested a scheme for its experimental verification. 
We found that the specific heat can be dramatically affected by \emph{superconducting} correlations. In particular, the strongest effects are predicted for intermediate or moderately-long diffusive SNS junctions, where both at low and high temperatures drastic deviations from the behavior of a normal metal are to be expected. Notably, the magnitude of the electronic specific heat can be finely and effectively controlled 
by changing the \emph{phase} difference between the superconductors. This  "active" phase-tuning of the thermodynamic properties of mesoscopic conductors may pave the way for novel applications, e.g., ultrasensitive  single-photon Josephson detectors based on proximity effect \cite{pjs}.

We acknowledge M. A. Shahzamanian for fruitful discussions, the Iranian Nanotechnology Initiative and the NanoSciERA "NanoFridge" project of the EU for partial financial support.


\begin{references}
\bibitem{interface}  We note that the presence of opaque NS interfaces weakens the proximity effect in the N region \cite{c4,hammer}, leading to a reduced influence on the electronic specific heat.
\end{references}
\end{document}